\documentstyle[subeqn]{article}
\begin{document}
\baselineskip = 18 pt

\begin{titlepage}
\begin{center}
\large {OPTICAL SOLITONS IN HIGHER ORDER NONLINEAR SCHR{\" O}DINGER EQUATION}
\end{center}
\vspace{2 cm}
\begin{center}
{\bf Sasanka Ghosh}$^*$ and {\bf Sudipta Nandy}$^{\dag}$\\
{\it Physics Department, Indian Institute of Technology, Guwahati}\\
{\it Panbazar, Guwahati - 781 001, India}\\
\end{center}
PACS numbers: 42.81.Dp, 02.30.Jr, 42.65.Tg, 42.79.Sz
\vspace{3 cm}
\begin{abstract}
We show the complete integrability and the existence of optical solitons of 
higher order nonlinear Schr{\" o}dinger equation by inverse scattering method 
for a wide range of values of coefficients. This is achieved first by invoking 
a novel connection between the integrability of a nonlinear evolution equation 
and the dimensions of a family of matrix Lax pairs. It is shown that Lax pairs 
of different dimensions lead to the same evolution equation only with the 
coefficients of the terms in different integer ratios. Optical solitons, thus 
obtained by inverse scattering method, have been found by solving an $n$ 
dimensional eigenvalue problem.  
\end{abstract}
\end{titlepage}
\newpage
\baselineskip = 24 pt

The propagation of optical solitons in an optical fibre draws a lot of attention 
in recent times because of its plausible applications in telecommunication. It 
is known that the propagation of optical pulses of very short duration in a 
fibre is described by the dynamics of a nonlinear evolution equation, namely 
higher order Schr{\" o}dinger equation (HNLS) [1-5]:
\begin{equation}
\partial_z E = i[\alpha_1 \partial_{\tau \tau} E + \alpha_2 \vert E \vert ^2 E] 
+ \epsilon [\alpha_3 \partial_{\tau \tau \tau} E 
+ \alpha_4 \partial_{\tau} (\vert E \vert ^2 E) 
+ \alpha_5 \partial_{\tau} (\vert E \vert ^2) E]
\label{I1}
\end{equation}
where, $E$ represents an envelop of electric fields, $z$ is the direction of 
propagation of optical pulse and $\tau $ is time. The terms on the r.h.s. of 
(\ref {I1}) describe respectively the effects of group velocity dispersion 
(GVD), self-phase modulation (SMP), third order dispersion (TOD), 
self-steepening (SS) and self-frequency shifting via stimulated Raman scattering 
(SRS).

If $\epsilon =0$, (\ref {I1}) reduces to nonlinear Schrodinger equation (NLS) and in
this case optical soliton can be produced due to a balance between GVD and SPM. 
But the optical soliton thus produced, cannot propagate over a long distance 
\cite {VI} with their shapes undistorted, due to the propagation loss, as they 
travel along the optical fibre. This propagation loss in the fibre, however, can 
be compensated by utilising Raman effect, which is achieved by transmitting a 
pump light wave simultaneously through the fibre \cite {VII}. It is interesting 
to note that the Raman process which takes care of propagation loss in the fibre, 
already exists within the spectrum of a soliton in HNLS equation (\ref {I1}) 
[1,2,8]. It is the last term, which is responsible for stimulated Raman 
scattering. No additional pump light wave is needed to compensate the 
propagation loss for the ultrashort optical pulses through the fibre. Moreover, 
for ultrashort pulses effects of higher order terms (last three terms) in 
(\ref {I1}) cannot be neglected. It is these reasons the study of the soliton 
solutions of HNLS equation becomes an important issue for the propagation of 
ultrashort pulses in the fibre. But, unfortunately the complete integrability of 
HNLS equation for arbitrary coefficients $\alpha $ is not well understood. It is 
our aim, in this paper, to study the Lax integrability and consequently to 
obtain soliton solutions of HNLS equation (\ref {I1}) by inverse scattering 
method (IST) for a wide range of values of the coefficients $\alpha $. 
Integrability of HNLS equation (1) by IST has so far been shown only for two 
fixed ratios of the coefficients [9,10].  

Now, in order to obtain soliton solutions of HNLS equation by IST, it is 
convenient to study a gauge equivalent nonlinear evolution equation,
\begin{equation}
\partial_{t}q(x,t) = \epsilon[\partial_{xxx}q(x,t) + \gamma_1 \vert q(x,t)
\vert^2 \partial_x q(x,t) + \gamma_2 \partial_x(\vert q(x,t)\vert^2)q(x,t)] 
\label{I2}
\end{equation}
which is related to the HNLS equation (\ref {I1}) explicitely through the 
following gauge transformation 
\begin{subequations}
\begin{eqnarray}
E(z,\tau) &=& exp[ -ipx + i\epsilon p^3t ] q(x,t)\\
\label{IIIa}
x &=& \tau + \alpha_1pz\\
\label{IIIb}
t &=& \alpha_3z
\label{IIIc}
\end{eqnarray}
\label{III}
\end{subequations}
with $p=\alpha_1/(3\epsilon \alpha_3)=\alpha_2/(\epsilon \alpha_4)$, $\gamma_1=
\alpha_4/\alpha_3$, and $\gamma_2=\alpha_5/\alpha_3$. The equation (\ref {I2}) 
is known as complex modified KdV (cmKdV) equation. Thus there is an one to one 
relation between the solitons of cmKdV equation and those of HNLS equation 
through the relation (\ref {III}a). Integrability of (\ref {I2}) is shown by 
many authors by various methods [9-13] only for two cases, when the coeffients 
of the terms on the r.h.s. are in the ratios (i) 1:6:0 and (ii) 1:6:3. Recently 
N solitary wave solutions of HNLS equation has been obtained for one parameter 
family of $\gamma $ parameters \cite {V}. But the Painleve analysis restricts 
the integrability of HNLS equation further to a subset of $\gamma $ parameters, 
when the terms in the r.h.s. of (\ref {I2}) are in the ratio 1:6:3. In this 
paper we show the complete integrability of HNLS equation by IST for all values of the $\gamma $ parameters.  
This is achieved by exploiting a novel connection between the dimensionality of 
the Lax pair and the integrability of HNLS equation. It is interesting to 
observe that the Lax pairs of different dimensions lead to the same equation of 
motion (\ref {I2}), but $\gamma_1$ and $\gamma_2$ in different ratios. 
Subsequently, we generalise the IST method for the $n$-dimensional Lax pair and 
obtain soliton solutions for HNLS equation for all integer valued ratios of the 
coefficients. 

In order to show the Lax integrability of HNLS equation vis a vis cmKdV 
equation, let us start with an $n$ dimensional Lax equations, 
\begin{subequations}
\begin{eqnarray}
\partial_x \Psi &=& {\bf U}(x,t,\lambda)\Psi \\     
\label{IVa}
\partial_t \Psi &=& {\bf V}(x,t,\lambda)\Psi
\label{IVb}
\end{eqnarray}
\label{IV}
\end{subequations}
where, ${\bf \Psi }(x,t)$ is an $n$ dimensional auxilliary field and the Lax 
operators ${\bf U}(x,t)$ and ${\bf V}(x,t)$ are $n\times n$ matrices of the 
form 
\begin{subequations}
\begin{eqnarray}
{\bf U} &=& -i \lambda {\bf \Sigma} + {\bf A} \\
{\bf V} &=& \epsilon {\bf A}_{xxx} + \epsilon ({\bf A}_x{\bf A} 
          -{\bf A}{\bf A}_x) + 2\epsilon {\bf A}^3 \\ \nonumber
     & &   -2i\epsilon \lambda {\bf \Sigma} ({\bf A}^2 -{\bf A}_x) 
        + 4\epsilon {\lambda}^2{\bf A}
        -4i\epsilon {\lambda}^3{\bf \Sigma}
\end{eqnarray}
\label{V}
\end{subequations}
In (\ref {V}) ${\bf \Sigma }$ is a c-no. diagonal matrix and the matrix ${\bf A}$ 
consists of dynamical fields, $q(x,t)$ and $q^*(x,t)$ only. The explicit form
of ${\bf \Sigma }$ and ${\bf A }$ may be given as 
\begin{subequations}
\begin{eqnarray}
{\bf \Sigma} &=& \sum_{i=1}^{n-1} e_{ii} - e_{nn} \\ 
\label{VIa}
{\bf A}(x,t) &=& \sum_{i=1}^{n-1} \alpha_i(x,t) e_{in}
 - \sum_{i=1}^{n-1}\alpha_i^*(x,t) e_{ni} 
\label{VIb}
\end{eqnarray}
\label{VI}
\end{subequations}
where, $e_{ij}$ is an $n\times n$ matrix whose only $(ij)$th. element is unity, 
the rest elements being zero and $\alpha_i(x,t)$ represent the dynamical fields. 
They may be chosen either as $q(x,t)$ or as $q^*(x,t)$. It is interesting to note 
that different choices of $\alpha_i$ lead to the same cmKdV equation (\ref {I2})  
only with the coefficients of the last two terms in different ratios. To obtain
the equation of motion for the dynamical fields $q$ and $q^*$, we first notice 
that 
\begin{equation}
{\bf \Sigma}^2 = {\bf 1}, \quad\quad 
{\bf \Sigma}{\bf A} + {\bf A}{\bf \Sigma} = 0
\label{IS}
\end{equation}
The equation of motion will then follow from the compatibility condition of 
(\ref {IV}), namely
\[ 
\partial_t {\bf U}(x,t) - \partial_x {\bf V}(x,t) + [{\bf U}(x,t) , {\bf V}(x,t)]
 = 0 
\]
and by using the relation (\ref{IS}) as
\begin{equation}
\partial_t{\bf A} = \epsilon \partial_{xxx}{\bf A} - 3\epsilon ({\bf A}^2 
(\partial_x{\bf A}) + (\partial_x{\bf A}){\bf A}^2)
\label{VIII}
\end{equation}
The above relation gives a nonlinear evolution equation for the matrix 
${\bf A}$. It is now straightforward to obtain evolution equations for $q(x,t)$ 
or $q^*(x,t)$, by choosing an explicit form of ${\bf A}$ (\ref {VIb}). We will 
see the coefficients $\gamma_1$ and $\gamma_2$ depend not only on the dimensions 
of the matrix ${\bf A}$, but also on specific choices of the dynamical fields 
$\alpha_i$ as $q$ or $q^*$. Let us consider some specific examples to clarify 
this point. 

If we consider a two dimensional Lax operator, ${\bf \Sigma}$ and ${\bf A}$ in
(\ref {VI}) would be of the form 
\begin{eqnarray*}
{\bf \Sigma} &=& e_{11} - e_{22}\\ 
&=& \left( \begin{array}{cc}
             1 & 0 \\
             0 & -1
\end{array} \right)    
\end{eqnarray*}
and
\begin{eqnarray*}
 {\bf A} &=& \alpha_1 e_{12} - \alpha_1^* e_{21}\\ 
&=& \left( \begin{array}{cc}
             0 & \alpha_1 \\
        -\alpha_1^* & 0
\end{array} \right)    
\end{eqnarray*}
and in this case only one dynamical field, $\alpha_1$ exists. Therefore, in 
two dimensional case the choice of Lax operator is unique, $\alpha_1=q$ and 
substituting ${\bf A}$ in (\ref {VIII}), it is found that the equation of motion 
reduces to the well known Hirota equation \cite {VIIIa}, 
\begin{equation}
\partial_t q = \epsilon \partial_{xxx} q + 6\epsilon\vert q\vert^2\partial_x q 
\label{IX}
\end{equation}  
where, the coefficients are in the ratio 1:6:0. 

However, for three dimensional Lax operator ${\bf \Sigma}$ and ${\bf A}$ are
of the form, 
\[
{\bf \Sigma} = e_{11} + e_{22} - e_{33}\quad\quad {\bf A} = \alpha_1e_{13} +
\alpha_2e_{23} - \alpha_1^*e_{31} - \alpha_2^*e_{32}
\]
We have, therefore, two possible choices of the Lax 
operators since two independent dynamical fields, $\alpha_1$ and $\alpha_2$ are
present in ${\bf A}$. In one case, we may choose both $\alpha_1$ and 
$\alpha_2$ as $q$ and consequently the equation of motion (\ref {VIII}) becomes 
\[
\partial_t q = \epsilon \partial_{xxx} q + 12\epsilon\vert q\vert^2\partial_x q 
\]
which reduces to Hirota equation (\ref {IX}) by appropriate scaling of the 
fields, $q$ and $q^*$. In another possible case, we may choose 
$\alpha_1$ as $q$, but $\alpha_2$ as $q^*$ and then both the Lax operator and 
the equation of motion become indentical to Sasa Satsuma case \cite {IX}:
\[  
\partial_t q = \epsilon \partial_{xxx} q + 6\epsilon\vert q\vert^2\partial_x q 
+ 3\epsilon \partial_x(\vert q\vert^2) q 
\]
where the coefficients are in the ratio 1:6:3. These two specific cases are 
already studied from various points of view. Moreover, no other ratios of the 
coefficients $\gamma$ emerge upto three dimensions from the family of Lax 
operators, considered here. We will see, however, some interesting consequences 
as we consider four dimensional Lax operators. It follows from (\ref {VI}) that 
for four dimensional Lax operators, ${\bf \Sigma}$ and ${\bf A}$ takes the form
\begin{eqnarray*}
{\bf \Sigma} &=& e_{11} + e_{22} + e_{33} - e_{44}\\ 
{\bf A} &=& \alpha_1e_{14} + \alpha_2e_{24} + \alpha_3e_{34} - \alpha_1^*e_{41} 
- \alpha_2^*e_{42} - \alpha_3^*e_{43}
\end{eqnarray*}
and eventually, three different choices of the Lax operators are manifest in 
this case.  One possibility, of course, may be considered when all the 
$\alpha_1$, $\alpha_2$ and $\alpha_3$ are chosen as $q$. The evolution equation 
(\ref {VIII}) then, once again, reduces to the known one, namely Hirota equation 
(\ref {IX}) by appropriate scaling of the fields. Interestingly, however, if we 
choose both $\alpha_1$ and $\alpha_2$ as $q$, but $\alpha_3$ as $q^*$, the 
evolution equation (\ref {VIII}) becomes
\begin{equation}  
\partial_t q = \epsilon \partial_{xxx} q + 12\epsilon\vert q\vert^2\partial_x q 
+ 3\epsilon \partial_x (\vert q\vert^2) q 
\end{equation}
where the coefficients are in the ratio 1:12:3. On the other hand, if we 
consider another possibility, where $\alpha_1$ is chosen as $q$, but both 
$\alpha_2$ and $\alpha_3$ as $q^*$, the evolution equation turns out to be 
\begin{equation}  
\partial_t q = \epsilon \partial_{xxx} q + 6\epsilon\vert q\vert^2\partial_x q 
+ 6\epsilon \partial_x(\vert q\vert^2) q, 
\end{equation}
the coefficients being in the ratio 1:6:6. The last two cases are definitely
new ones, which will be shown to be integrable. It is now clear that as we go to 
higher and higher diemnsions, more and more new ratios of the coefficients 
appear for which cmKdV equation will be integrable. If fact, in general for an 
$n$ dimensional Lax operator, as is evident from (\ref {VI}), $(n-1)$ number of
dynamical fields, $\alpha_i$ exist. We may, therefore, identify $l$ number of 
$\alpha$'s as $q$ and the rest $(n-l-1)$ ones as $q^*$. Consequently, the 
evolution equation (\ref {VIII}) yields as 
\begin{equation}  
\partial_t q = \epsilon \partial_{xxx} q + 6l\epsilon\vert q\vert^2\partial_x q 
+ 3(n-l-1)\epsilon \partial_x(\vert q\vert^2) q, 
\end{equation}
which is nothing but cmKdV equation, where the coefficients are in the ratio 
1:6$l$:3$(n-l-1)$, i.e
\begin{equation}
\gamma_1 = 6l, \quad\quad\quad \gamma_2 = 3(n -l- 1).
\label{Xa}
\end{equation}
cmKdV equation is, thus, Lax integrable for all possible integer values of 
coefficients, $\gamma_1$ and $\gamma_2$, which can be domonstrated by 
appropriately scaling the dynamical fields, $q$ and $q^*$ in each case. It is 
evident now, for a given $n$ dimensional matrix, $(n-1)$ Lax pairs can be 
constructed and each Lax pair of a given dimension, leads to $(n-1)$ different 
ratios of the coefficients. This is a significant achievement over the previous 
works. The existence of Lax pair, although, is a storng evidence for the 
integrability of cmKdV equation, it remains to show that the family of Lax pairs 
(\ref {V}) admit soliton solutions through IST. 

In order to obtain soliton solutions through IST in our case, we have 
generalised the $3\times 3$ AKNS type eigenvalue problem [10,14] to $n\times n$ 
eigenvalue problem. Notice that asymptotically the auxiliary field, 
${\bf \Psi }(x,t)$ obeys a simple relation, viz.
\begin{equation}
{\bf \Psi}(x,t) = exp[\{-i\lambda x - 4i\epsilon \lambda^3t\}{\bf \Sigma}],
\label{XI}
\end{equation}
for the whole family of the Lax pairs (\ref {V}) and it depends only on the 
dimensions of the matrix Lax pair. Consequently the scattering data matrix, 
which, by definition, connects the Jost functions, 
$\phi^{(i)}(x,\lambda )\vert_{x=-\infty}$  to the Jost functions, 
$\psi^{(i)}(x,\lambda )\vert_{x=\infty}$, for $i=1,2,\cdots,n$, [10,15] 
evolves with time in some universal form. Finally, solving $n$ coupled Gelfand 
Levitan Marchenko equations one soliton solution may be expressed in a simple 
form
\begin{equation}
q(x,t) = (\eta/{\sqrt(n-1)})sech A(x,t) exp(iB(x,t))
\label{XII}
\end{equation}
with 
\begin{eqnarray*}
A(x,t) &=& \eta x - \epsilon (\eta^3 - 3\xi^2 \eta )t - \gamma 
-{1\over 2} ln(n-1)\\
B(x,t) &=& \xi x + \epsilon (\xi^3 - 3\xi \eta^2)t + \delta 
\end{eqnarray*}
where $\gamma$ and $\delta$ are determined by the initial conditions and we 
assume that the simple pole is situated in the upper half plane at 
$\lambda_1={1\over 2}(-\xi+i\eta)$ for one soliton solution. A detailed 
calculation of soliton solutions by IST method for an $n$ dimensional
eigenvalue problem is considered in \cite {XIII} for a more general case than
HNLS equation and will be published elsewhere. It is straightforward to obtain 
optical soliton in terms of the original parameters by substituting the relations 
(\ref {III}) in (\ref {XII}) and it turns out to be
\begin{equation}
E(z,\tau) = (\eta/{\sqrt(n-1)})sech {\tilde A}(z,\tau) exp(i{\tilde B}(z,\tau))
\label{XIII}
\end{equation}
with 
\begin{eqnarray*}
{\tilde A}(z,\tau) &=& \eta \tau - [\epsilon \alpha_3(\eta^3 - 3\xi^2 \eta )
 - {\alpha_1^2\over{3\epsilon \alpha_3}}\eta ]z - \gamma - {1\over 2} ln(n-1)\\
{\tilde B}(z,\tau) &=& (\xi - {\alpha_1\over{3\epsilon \alpha_3}} \tau 
+ [\epsilon \alpha_3(\xi^2 - 3 \eta^2 
+ {\alpha_1^2\over {3\epsilon^2\alpha_3^2}})\xi 
- {2\alpha_1^3\over{27\epsilon^2 \alpha_3^2}}]z + \delta 
\end{eqnarray*}
It is interesting to observe from (\ref {XIII}) that the envelop of one soliton 
solution admits a simple {\it sech} type shape, which can be easily produced by
a mode-locked laser. Moreover, the intensity, $I_s=\eta^2/(n-1)$ and the width, 
$\Gamma_s=\eta^{-1}$ for one soliton in (\ref {XIII}) are related by the 
expression $I_s\Gamma_s^2=1/(n-1)$. In terms of $\gamma$ parameters, given in 
(\ref {Xa}), the expression $I_s\Gamma_s^2$ becomes 
$I_s\Gamma_s^2=6/\{3\gamma_1+2(\gamma_2-\gamma_1)\}$, which is in full 
agreement with the result of \cite {V}. 

To conclude, the exact integrability of the cmKdV equation is shown for the 
integer valued ratios of the coefficients by establishing an intreguing 
relationship of the coefficients $\gamma$ with the dimensionality of the Lax 
pairs. The family of Lax pairs, so obtained, is shown to admit soliton solutions. 
One soliton solutions are found by solving an $n\times n$ eigenvalue problem 
through IST. For one soliton, the relation between the intensity and the width
turns out to be simple and depends only on the diemnsions of the Lax pairs. 

S.N. would like to thank CSIR, Govt. of India for financial support and for 
the award of Junior Research Fellowship.

Electronic address : $^*$sasanka@iitg.ernet.in, $^{\dag}$sudipta@iitg.ernet.in


\begin{thebibliography}{99}
\bibitem{I} Y. Kodama and A. Hasegawa, IEEE J. Quantum Electron. QE-23,5610 
(1987). 
\bibitem{II} F.M. Mitschke and L.F. Mollenauer, Opt. Lett. 11, 657 (1986). 
\bibitem{III} A. Hasegawa, Optical Solitons in Fibres, (Springer, Heidelberg, 
1989).
\bibitem{IV} G.P. Agrawal, Nonlinear Fibre Optics, (Academic Press, NY, 1989).
\bibitem{V} M. Gedalin, T.C. Scott and Y.B. Band, Phys. Rev. Lett. 78, 448 
(1997).
\bibitem{VI} L.F. Mollenauer, R.H. Stolen and J.P. Gordon, Phys. Rev. Lett 45, 
1095 (1980).
\bibitem{VII} A. Hasegawa, Appl. Opt. 23, 3302 (1984).
\bibitem{VIII} J.P. Gordon, Opt. Lett. 11, 662 (1986).
\bibitem{VIIIa} R. Hirota, J. Math. Phys. 14, 805 (1973). 
\bibitem{IX} N. Sasa and J. Satsuma, J. Phys. Soc. Jpn. {\bf 60}, 409 (1991).
\bibitem{X} K. Porsezein, M. Daniel and M. Lakshmanan, Proc. of the Int. Conf.
on Nonlinear Evolution Equations and Dynamical Systems, Eds. V.G. Makhankov, 
I. Puzynin and O. Pashaev 436 (World Sc., Singapore).
\bibitem{XI} K. Porsezian and K. Nakkeeran, Phys. Rev. Lett. 76, 3955 (1996).
\bibitem{XII} S. Ghosh, A. Kundu and S. Nandy, Soliton Solutions, Liouville 
Integrability and Gauge Equivalence of Sasa Satsume Equation, J. Math. Phys. (to
be published).
\bibitem{XIIa} S.V. Manakov, Sov. Phys.-JETP 38, 248 (1974).
\bibitem{XIII} S. Ghosh and S. Nandy, Inverse Scattering Method and Vector
Higher Order Nonlinear Schr{\" o}dinger Equation (under preparation).
\end{thebibliography}
\end{document}